\documentclass[twocolumn]{aastex631}
\usepackage{amsmath}

\newcommand\aastex{AAS\TeX}

\begin{document}

\title{Template \aastex Article with Examples: 
v6.3.1\footnote{Released on March, 1st, 2021}}

\title{Distribution and evolution of Li abundance in red clump stars can be explained by the internal gravity waves}

\shortauthors{Li et al.}

\email{lixuefeng@ynao.ac.cn}

\author[0000-0001-9291-4261]{Xue-Feng Li}
\affiliation{Yunnan Observatories, Chinese Academy of Sciences, P.O.Box110, Kunming 650216, China}

\affiliation{University of Chinese Academy of Sciences, Beijing 100049, China}

\author[0000-0002-0349-7839]{Jian-Rong Shi}
\affiliation{CAS Key Laboratory of Optical Astronomy, National Astronomical Observatories, Beijing 100101, China}
\affiliation{School of Astronomy and Space Science, University of Chinese Academy of Sciences, Beijing 100049, China}

\author[0000-0002-1424-3164]{Yan Li}
\affiliation{Yunnan Observatories, Chinese Academy of Sciences, P.O.Box110, Kunming 650216, China}
\affiliation{University of Chinese Academy of Sciences, Beijing 100049, China}
\affiliation{Key Laboratory for Structure and Evolution of Celestial Objects, Chinese   Academy of Sciences, P.O.Box110, Kunming 650216, China}
\affiliation{Center for Astronomical Mega-Science, Chinese Academy of Sciences, Beijing 100012, China}

\author[0000-0002-8609-3599]{Hong-Liang Yan}
\affiliation{CAS Key Laboratory of Optical Astronomy, National Astronomical Observatories, Beijing 100101, China}
\affiliation{School of Astronomy and Space Science, University of Chinese Academy of Sciences, Beijing 100049, China}
\affiliation{Institute for Frontiers in Astronomy and Astrophysics, Beijing Normal University,  Beijing 102206, China}

\author[0000-0002-2510-6931]{Jing-Hua Zhang}
\affiliation{CAS Key Laboratory of Optical Astronomy, National Astronomical Observatories, Beijing 100101, China}

 
\begin{abstract}
The study of Li phenomena in red clump (RC) stars can give us a deeper understanding of the structure and evolution of stars. \citet{2022ApJ...933...58C} explained the RC Li abundance distributions naturally using only standard post main sequence (MS) Li evolution models when the distribution of progenitor masses and the depletion of Li during the MS observed in MS stars were considered, thus neither extra Li depletion nor Li creation mechanism is required. Nevertheless, it is interesting to consider the effects of mixing caused by some extra mechanisms.
By constructing different models, we find that the mixing caused by internal gravity waves can explain the observed Li abundances of RC stars with low mass progenitors. To explain that, we rely on the extra mixing induced by internal gravity waves that are excited at the bottom of the convective envelope at the red giant branch (RGB) stage. During the RGB stage, introducing the internal gravity waves can improve the diffusion coefficient and strengthen the mixing effect. The effective enrichment of Li occurs at the late RGB stage and requires the diffusion coefficient of H-burning shell to reach $\rm \sim 10^{8}\,cm^{2}\,s^{-1}$. Our models predict that the Li abundance decreases from $\rm \sim 1.5\,dex$ to $\rm \sim 0.0\,dex$ at the end of core He-burning stage, thereby revealing the $\sim 99\%$ of the observed Li abundance distribution. The thermohaline mixing regulates the Li abundance of RGB stars, which combines with the internal gravity waves can explain the Li abundances of most giants.
\end{abstract}


\section{Introduction} \label{sec:intro}

The abundance of lithium (Li) has a comprehensive performance of the complex thermonuclear reaction process and element mixing process inside stars. The study of Li abundance on the surface of stars can test the theory of stellar structure and evolution, and explore some key physical processes inside stars.

The definition of the Li abundance is: $\rm A_{Li}=log(N_{Li}/{N_{ H}})+12$, where $\rm N_{Li}$ and $\rm N_{H}$ are the atomic number densities of $\rm Li$ and $\rm H$, respectively. \citet{2021ApJ...914..116G} determined the Li abundances of 165,479 stars from the Large Sky Area Multi-Object Fiber Spectroscopic Telescope \citep[LAMOST,][]{2012RAA....12.1197C, 2022Innov...300224Y} medium-resolution survey. The Li abundances of these stars are mainly distributed between $\rm 0.5\,dex$ and $\rm 3.0\,dex$, and there are two obvious peaks around 1.0$\rm \,dex$ and 2.6$\rm \,dex$, which are dominated by giants and hot dwarfs, respectively \citep{2021ApJ...914..116G}. Based on the GALAH \citep{2017MNRAS.465.3203M, 2018MNRAS.478.4513B} and $\textsl{K2}$-HERMES \citep{2018AJ....155...84W, 2019MNRAS.490.5335S} surveys, the Li abundances of 109,340 giants analysed by \citet{2021MNRAS.505.5340M} are distributed between $\rm -1.0\,dex$ and 4.0$\rm \,dex$. It was also found that the Li abundances of a fraction of giant stars are higher than $1.5$$\rm \,dex$, which is a traditional definition of the so-called Li-rich giant stars. Although some recent researches show that this threshold is flawed \citep[e.g.,][]{2022ApJ...933...58C, 2022MNRAS.513.5387S}, we still follow this threshold  in this work. The percentage of Li-rich giant stars is around $\rm \sim 1\%$  \citep{1989ApJS...71..293B, 2019ApJS..245...33G, 2021MNRAS.505.5340M}.

The red clump (RC) stars are in the core He-burning stage, which are always crowded with red giant branch (RGB) bump stars in their H-R diagrams \citep{2016ARA&A..54...95G}. Thanks to the  asteroseismology, we can separate the RC stars from the RGB ones \citep{2011Natur.471..608B}. A considerable number of Li-rich giant stars are recognized as RC stars \citep{2014ApJ...784L..16S, 2018ApJ...858L..22B, 2019ApJ...880..125C, 2019ApJ...878L..21S, 2019ApJ...877..104Z, 2021MNRAS.505.5340M, 2021NatAs...5...86Y}. Based on the astroseismology and spectroscopy, \citet{2021NatAs...5...86Y} provided a clear evidence that most of low-mass Li-rich giant stars are RC stars.


Recently, \citet{2020NatAs...4.1059K} collected a sample of more than 26,000 giant stars from GALAH DR2 \citep{2018MNRAS.478.4513B}, and further selected the densest part in the RC region on the HR diagram, and finally obtained 9,284 RC stars. $\rm A_{Li}$ in these RC stars presents a large-scale distribution from $\rm -0.9\,dex$ to $\rm 3.7\,dex$. They found that about $97\%$ of RC stars have $\rm A_{\rm Li}$ within $0.0-1.5$$\rm \,dex$, which forms a dense $\rm A_{\rm Li}$ distribution area. In a sample of 944 RC stars obtained by \citet{2021ApJ...919L...3Z} through asteroseismology analysis, similar Li abundance distribution is found as that by \citet{2020NatAs...4.1059K}, except that almost all RC stars have a $\rm A_{Li}>0.5\,dex$ in their sample.

Many scenarios have been proposed to explain the observed Li abundance distribution of RC stars. \citet{2022ApJ...933...58C} explained the RC Li abundance distributions naturally using only standard post main sequence (MS) Li evolution models when the distribution of progenitor masses and the depletion of Li during the MS  were  considered. 
Therefore, neither extra Li depletion nor Li creation mechanism is required. Nevertheless, some extra physical mechanisms have been discussed.
$\rm A_{\rm Li}$ can increase by about one order of magnitude when \citet{2021MNRAS.503.2746M} introduced the neutrino magnetic moment, and the final distribution was between $\rm -0.5\,dex$ and 0.5$\rm \,dex$.  \citet{2020ApJ...901L..18S} found, for the first time, that the mixing process induced by internal gravity waves that generated by convection during the first He flash can enrich Li and interpret the Li enhancement of RC stars.

\citet{1955ApJ...121..144C} and \citet{1971ApJ...164..111C} proposed a mechanism to explain the Li problem in giants, i.e., Be produced by H burning in the hotter region inside stars is transported to the cooler outer region, and decays into Li. The operation of Cameron-Fowler mechanism requires an appropriate mixing process between the convective envelope and the H-burning shell, and the asymptotic giant branch (AGB) stars have hot bottom burning so that the bottom of the convective envelope can reach the H-burning shell. But for RC stars, the envelope is difficult to reach the H-burning shell, hence, extra mixing should be operated in this region. \citet{2020ApJ...901L..18S} first considered the mixing effect caused by internal gravity waves that is excited by the convection zone inside the star during the He flash, and obtained a considerable order of magnitude distribution of the diffusion coefficient. Inspired by this idea \citep{2020ApJ...901L..18S}, we consider the mixing effect caused by internal gravity waves induced by another convection zone, i.e., the convective envelope, at giant stage. 

Internal gravity waves excited at the bottom of the convective envelope can propagate inward, and  will be reflected at a place where the mass fraction of H reduces rapidly. Therefore, a resonant cavity will be formed. Meanwhile, standing waves are also formed within it, which will lead to a strong mixing process that can induce the enrichment of Li. The resonant cavity region of internal gravity waves just provides the operating area for the Cameron-Fowler mechanism. 

In this paper, we apply the mixing effect of internal gravity waves to explain the Li abundance distribution and evolution of RC stars with low mass progenitors. In $\rm Section\,\ref{sec:Data}$, we simply analyze our data samples. We introduce the tools and some basic settings of the models and discuss the analytic expressions of the diffusion coefficient in $\rm Section\, \ref{sec:Method}$. In $\rm Section\,\ref{sec:Model}$, we show the deficiencies of the standard model, and consider several extra mixing. $\rm Section\,\ref{sec:Discussion}$ mainly discuss the model for stars with masses greater than $1.5M_{\odot}$. Finally, conclusions are presented in $\rm Section\, \ref{sec:Conclusion}$.

\section{Data} \label{sec:Data}
\begin{figure*}[t]
	\centering
	\plotone{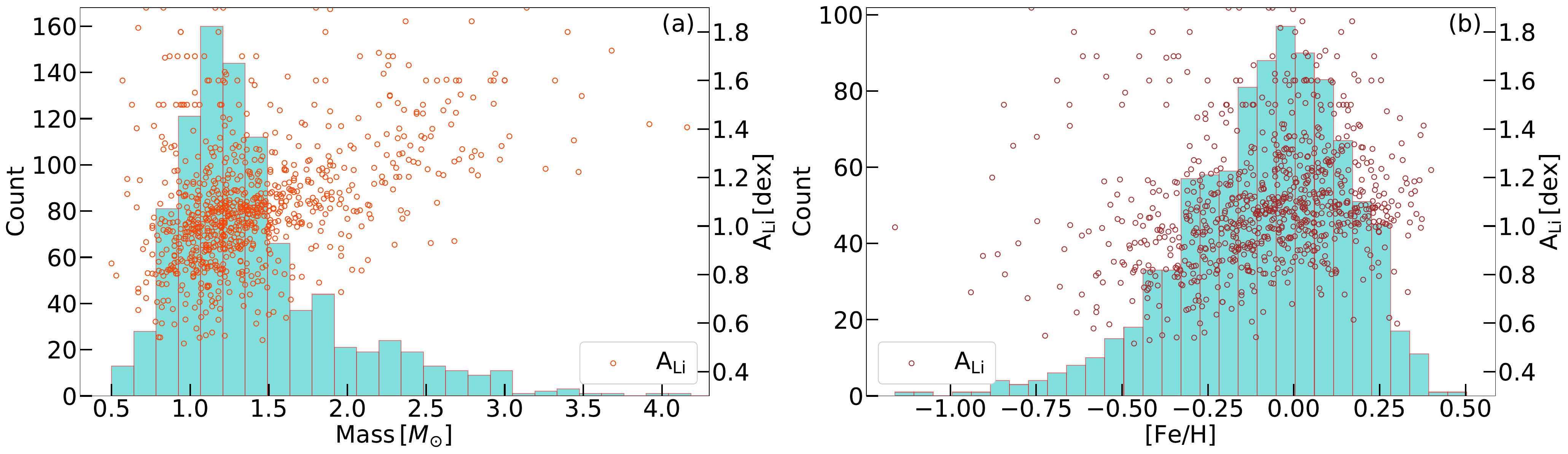}
	\caption{The Li abundance vs mass and [Fe/H] of RC stars. The left Y-axis is the number of stars, and the right Y-axis is the Li abundances of these stars. This sample comes from \citet{2021ApJ...919L...3Z}.\label{fig:f1}}
\end{figure*}

Our data sample includes 944 RC stars selected by \citet{2019ApJS..245...33G}, \citet{2021ApJ...914..116G}, \citet{2021NatAs...5...86Y} and \citet{2021ApJ...919L...3Z} from the LAMOST surveys. These stars are identified using asteroseismology analysis, and the typical error of Li abundance is $\rm 0.2\,dex$. $\rm Figure\,\ref{fig:f1}$ shows the distribution of Li abundance as functions of mass and [Fe/H] for this sample. It can be seen that the mass and [Fe/H] of these stars are mainly concentrated between $\sim 0.8-1.8M_{\odot}$ and $\rm -0.5-+0.25\,dex$, and the peaks are around $1.2M_{\odot}$ and $\rm -0.05\,dex$, respectively.

\section{Method} \label{sec:Method}
\subsection{Inputs} \label{sec:input}
We evolve our models by the Modules for Experiments in Stellar Astrophysics (MESA, release: 11701 \citep{2011ApJS..192....3P, 2013ApJS..208....4P, 2015ApJS..220...15P, 2018ApJS..234...34P, 2019ApJS..243...10P}). The equation of state tables from \citet{2002ApJ...576.1064R} are used, and the OPAL opacity tables from \citet{1993ApJ...412..752I, 1996ApJ...464..943I} are adopted. The nuclear reaction network we selected is \textbf{\textit{pp\_extras.net}}, which involves a total of 12 elements, mainly including $\rm ^{3}He$, $\rm ^{4}He$, $\rm ^{7}Be$, $\rm ^{7}Li$, and $\rm ^{8}B$. These elements are involved in the Cameron-Flower Be transfer mechanism. The treatment of convection is based on \citet{1968pss..book.....C}. The chemical composition of our models is from GS98 \citep{1998SSRv...85..161G}. 

We proceed from two perspectives, the first from a model perspective, and the second utilizing the observed Li abundance as the initial input and validating our models. In our models, we start to construct the model from the pre-MS, let the stars evolve naturally, and consider additional physical processes at a specific time. In the MESA code, the initial $\rm A_{Li}$ varies with initial Z ($\rm (A_{Li})_{initial}=3.4+[Fe/H]$), and the corresponding initial $\rm A_{\rm Li}$ is $\rm \sim 3.4\,dex$ at the solar metallicity \citep{1998SSRv...85..161G}. We need to start from the PMS to build the model, so we use the meteoritic value from \citet{1998SSRv...85..161G} as the initial Li abundance.
We add the expression of diffusion coefficient for the mixing process induced by internal gravity waves to the \textbf{\textit{run\_star\_extras.f}} of the MESA code, which is described in detail below.

\subsection{Diffusion Coefficient} \label{sec:diffusion}

The key to the Li enhancement of RC stars is a proper diffusion coefficient distribution at the regions where $\rm ^{7}Be$ and $\rm ^{7}Li$ are produced, which ensures that $\rm ^{7}Be$ can be efficiently transported to the convective envelope.

\citet{1994A&A...281..421M}, \citet{1996A&A...305..513M}, and \citet{2000A&A...354..943M} systematically studied the effect of mixing process caused by internal gravity waves on Li and Be abundances in solar-like MS stars. \citet{1994A&A...281..421M} and \citet{2000A&A...354..943M}, respectively, considered perturbations that excited the internal gravity waves at the convective boundary based on diverse convection treatment methods, and obtained approximate results.

Inspired by the work of \citet{1981ApJ...248..751P} and \citet{1991A&A...252..179Z}, \citet{1994A&A...281..421M} deduced the expression of the diffusion coefficient, which is given by:

\begin{equation}\label{equ:e1}
\begin{split}
D_{\rm mix} & = A^{2} \frac{3}{n+7} \frac{3}{n+5} \times \\
& D_{\rm th}^{2} \frac{N^{2}}{2\pi} (\frac{\rho_{\rm b}}{\rho})^{2} (\frac{r_{\rm b}}{r})^{12} V^{8n-3} l^{1-2n} f^{-2n},
\end{split}
\end{equation}
here, $A$ is a numerical factor. $\rho$, $r$ and $N$ are the density, radius and buoyancy frequency at each depth, respectively. $V$ and $l$ respectively represent the  velocity and scale of eddy. These quantities with subscript b represent the value of the corresponding quantity at the lower boundary of convective envelope. $n$ refers to the ripple morphology chosen for the perturbation, and \citet{1996A&A...305..513M} suggested $n = 1$.

For the thermal diffusivity $D_{\rm th}$:

\begin{equation}\label{equ:e2}
D_{\rm th}=\frac{16\sigma T^{3}}{3 c_{\rm p} \kappa \rho^{2}},
\end{equation}
where, $\sigma$ is the Stefan-Boltzmann constant, $T$ is temperature, $c_{\rm p}$ is the specific heat at constant pressure and $\rm \kappa$ is opacity.

$ f$ refers to the damping factor, and depends on $r$:

\begin{equation}\label{equ:e3}
f=| \int_{r_{\rm b}}^{r} D_{\rm th} N^{3} (\frac{r_{\rm b}}{r})^{3} dr |.
\end{equation}

\citet{2020ApJ...901L..18S} employed the internal gravity waves generated by the convective turbulence inside stars during the first He flash and chose $A = 1$. The selected characteristic eddy scale, and the eddy velocity were $\rm 10^{9}\,cm$ and $\rm10^{6}\,cm\,s^{-1}$, respectively. A suitable range of diffusion coefficient for Li enhancement is obtained, and the scope of diffusion coefficient is about $\rm 10^{8}-10^{11}\,cm^{2}\,s^{-1}$. Furthermore, the distribution of the diffusion coefficient shows an outward decreasing trend. The corresponding form is:

\begin{equation}\label{equ:e4}
D_{\rm mix}=\frac{A}{2\pi}[D_{\rm th} N (\frac{\rho_{\rm b}}{\rho})(\frac{r_{\rm b}}{r})^{6}]^{2} V^{5} l^{-1} f^{-2}.
\end{equation}

We study the internal gravity waves excited at the bottom of the convective envelope. \citet{2000A&A...354..943M} used the model of convective transport by plumes that described by \citet{1995A&A...296..127R} for the solar-like stars, which avoids to introduce temporary parameters in the convective transport by the plumes models. The corresponding diffusion coefficient is expressed as:

\begin{equation}\label{equ:e5}
D_{\rm mix}=A [D_{\rm th} N (\frac{\rho_{\rm b}}{\rho})(\frac{r_{\rm b}}{r})^{6}]^{2} V^{8n-3} l^{1-2n} f^{-2n},
\end{equation}
where $n=\frac{2}{3}$.

\subsection{Velocity and Scale} \label{sec:avl}

We use analytical expressions to calculate the velocity and scale of eddy. $V$ can be obtained by using the mixing length theory:

\begin{equation}\label{equ:e6}
V=(\frac{\phi L_{\rm b}}{4\pi r_{\rm b}^2\rho_{\rm b}})^{1/3},
\end{equation}
where, $\rm \phi$ is a constant with a standard value of 0.1. The luminosity at the convective boundary is expressed as $L_{\rm b}$.

$l$ is the characteristic horizontal eddy scale. It is not suitable to take the scale of eddy as a constant of about $\rm 10^{9}\,cm$ for different evolutionary phases, so we reevaluate the scale of eddy and improve its calculation. \citet{2012ApJ...756...37L} proposed a $k-\omega$ model to describe turbulently thermal convection and gave the relationship between the velocity and the radius of eddy:

\begin{equation}\label{equ:e7}
R=\sqrt{\frac{\lambda}{\rho c_{\rm p} \Omega }},
\end{equation}
where $R$ and $\rm \Omega$ are the radius and angular velocity of eddy, respectively. The expression of $\lambda$ is:

\begin{equation}\label{equ:e8}
\lambda=\frac{16\sigma T^{3}}{3 \kappa \rho}.
\end{equation}

The velocity of eddy can be expressed as:

\begin{equation}\label{equ:e9}
V=R\Omega.
\end{equation}

Combining Eq.(\ref{equ:e7}), Eq.(\ref{equ:e8}) and Eq.(\ref{equ:e9}),  we obtain the following results:

\begin{equation}\label{equ:e10}
l=2R=\frac{2D_{\rm thb}}{V},
\end{equation}
where $D_{\rm thb}$ is the thermal diffusivity at the bottom of the convective envelope. Finally, we get:

\begin{equation}\label{equ:e11}
D_{\rm mix}=A(2D_{\rm thb})^{-\frac{1}{3}} [D_{\rm th}N (\frac{\rho_{\rm b}}{\rho})(\frac{r_{\rm b}}{r})^{6}]^{2}V^{\frac{8}{3}} f^{-\frac{4}{3}}.
\end{equation}

\subsection{Mixing area and action time}

How to incorporate the diffusion coefficient from the internal gravity waves into the models of giant stars? Firstly, we specify the propagation region of the internal gravity waves. We analyze the distribution of H mass fraction and the location of H-burning shell, and search for its jump point as the lower boundary. The H mass fraction of jump point is $\sim$ 0.65 for stars with solar metallicity. The temperature of this area is generally about $4\times10^{7}\,$K, which is a little bit lower than the location where the maximum of Be is produced. Since the internal gravity waves cannot propagate within the convective envelope, we regard the bottom of convective envelope as the upper boundary where the internal gravity waves excited. Thus Eq.(\ref{equ:e11}) is employed within the region between these specific points. 

Secondly, we consider the action time of mixing process induced by the internal gravity waves. Eq.(\ref{equ:e11}) is valid under the frequency conditions $N^{2}>>\omega^{2}$, where $\omega^{2}$ is approximately $\rm 10^{-12}\,s^{-2}$. $N^{2}$ between the H-burning shell and convective envelope increases from the outside to inside, so its minimum value is at the bottom of convective envelope. It is found that the minimum value of $N^{2}$ is approximately $\rm 10^{-10}\,s^{-2}$ when the mass of $\rm He$ core increases to about $0.30M_{\rm \odot}$, which is much greater than $\rm 10^{-12}\,s^{-2}$. It needs to be pointed out that not all of the region between the H-burning shell and the convective envelope meet the requirement of $N^{2}>>\omega^{2}$ when the mass of He core is less than $0.30M_{\rm \odot}$. In addition,  when the mass of He core is less than $0.30M_{\odot}$, the subsequent phases are likely to fail to meet the frequency condition, although the internal gravity waves are considered.

We use $\rm M_{He}$ to represent the mass of the He core. The following results are all obtained by considering the mixing effect induced by internal gravity waves in the period when $\rm M_{He}$ increases to $0.30M_{\rm \odot}$.

\section{The standard model and extra mixing}\label{sec:Model}
At present, the research of stellar model in regard to the Li abundances of RC stars shows a positive scene. 
Several other mechanisms have been proposed \citep{2020ApJ...901L..18S, 2021MNRAS.503.2746M}. Although they can explain the distribution to a certain extent, they have difficulty on explaining the evolution of Li abundance. Considered the distribution of progenitor masses and the deplete of Li during the MS observed in MS stars, recently, \citet{2022ApJ...933...58C} explained the RC Li abundance distributions naturally using standard post-MS Li evolution models. 
Thus, neither extra Li depletion nor Li creation mechanism is required. Nevertheless, it is interesting to consider the effects of mixing caused by some extra mechanisms (e.g. internal gravity waves), which is the focus of this study.


For this point, we start from the standard convection models, and take several extra mixing into consideration.
\subsection{Standard convection models}
\begin{figure*}[t]
	\centering
	\plotone{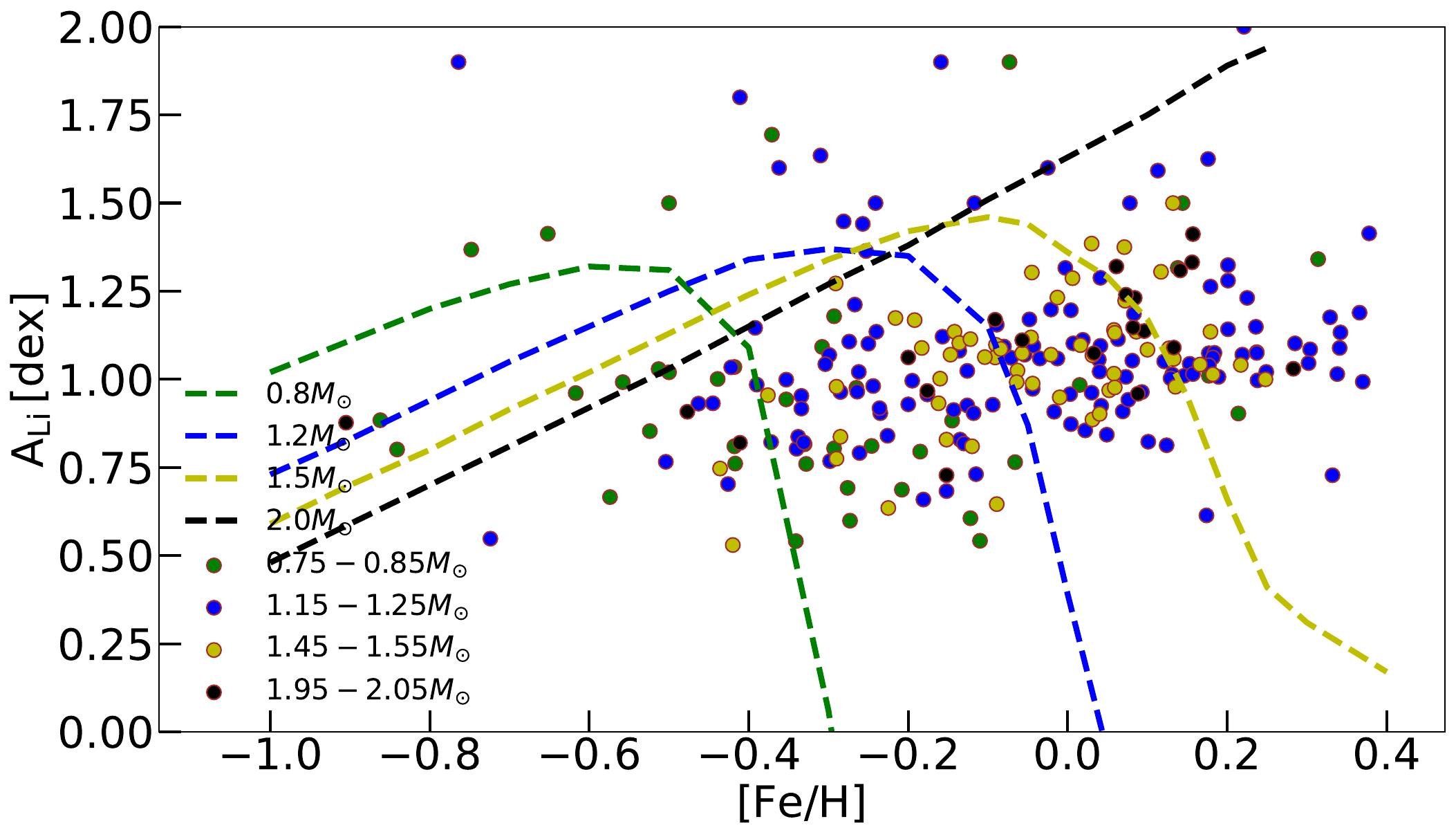}
	\caption{The distribution of Li abundances as a function of metallicity for the observed and model predicted results. The dashed lines are the model results. The circles are RC samples from \citet{2021ApJ...919L...3Z}, with different colors representing specific mass range. The discrete points in the subsequent figures without special instructions are from this sample.\label{fig:f2}}
\end{figure*}

Similar to  \citet{2022ApJ...933...58C}, we firstly considered models in which only convection effects involved, which we referred as standard convection model.
$\rm Table\,\ref{tab:t1}$ lists the Li abundances at different stages for our standard convection models, from which similar results to \citet{2022ApJ...933...58C} can be found. To validate the standard convection models, we take the effect of metallicity into account.

$\rm Figure\,\ref{fig:f2}$ shows the difference between the observed Li abundances of RC stars and the predicted results of standard models.  
For our convection models, when the stars of masses are less than $1.5M_{\odot}$, they are contrary to the observed results especially with higher [Fe/H], and the smaller the mass is, the greater the deviation is. For $2.0M_{\odot}$ stars, they show similar distribution characteristics as the observations, i.e., the Li abundances increase with [Fe/H]. But, the model results are overall $\sim \rm 0.5\,dex$ higher than those of the observations. 
Therefore, the results given by our convection models cannot fully explain the Li problem of RC stars. Our convection models do not take into account the Li depletion during  the MS. \citet{2022ApJ...933...58C} were able to explain the distribution of RC Li abundances when convection models are considered together with the distribution of RC masses (and especially the progenitor masses) and the distribution of Li depletion actually observed in MS stars.

We note there is increasing trend of $\rm A_{Li}$ with metallicities. The increasing slope is reminiscent of Galactic Li production\citep{2019MNRAS.489.3539G, 2021A&ARv..29....5M}, which implies similar  Li depletions at different metallicities.

\begin{deluxetable*}{ccccccrr}
	\tablenum{1}
	\tablecaption{Li abundance in different models\label{tab:t1}}
	\tablehead{\colhead{Model} & \colhead{Initial mass} & \colhead{Initial Z} & \colhead{Initial $\rm A_{Li}$} & \colhead{ZAMS} & \colhead{MS turnoff} & \colhead{RGB tip} & \colhead{The core He-burning phase}\\
		\colhead{} & \colhead{($M_{\odot}$)} & \colhead{} & \colhead{$\rm (dex)$} & \colhead{$\rm (dex)$} & \colhead{$\rm (dex)$} & \colhead{$\rm (dex)$} & \colhead{(dex)}}		
	\startdata
	& $2.0$ & $0.02$ & 3.40 & 3.39 & 3.39 &1.63 & $1.63$\\
	only	& $1.5$ & $0.02$ & 3.40 & 3.35 & 3.35 &1.36 & $1.36$\\
	convection & $1.2$ & $0.02$ & 3.40 & 3.14 & 3.12 &$0.40$ & $0.40$\\
	& $1.0$ & $0.02$ & 3.40 & 2.66 & 2.46 &$-2.08$ & $-2.08$\\
	\hline
	& $2.0$ & $0.02$ & 3.40 & 3.39 & 3.39 &1.63 & $1.63$\\
	convection +	& $1.5$ & $0.02$ & 3.40 & 3.30 & 3.30 &0.76 & $0.76$\\
	overshooting\tablenotemark{a}  	& $1.2$ & $0.02$ & 3.40 & 2.80 & 2.67 &$-1.96$ & $-1.96$\\
	& $1.0$ & $0.02$ & 3.40 & 1.68 & 1.19 &$-6.10$ & $-6.10$\\
	\hline
	& $2.0$ & $0.02$ & 3.40 & 3.39 & 3.39 &1.59 & $+1.59-+1.54$\\
	convection +		& $1.5$ & $0.02$ & 3.40 & 3.35 & 3.35 &1.26 & $+1.26-+0.93$\\
	thermohaline mixing\tablenotemark{b}	& $1.2$ & $0.02$ & 3.40 & 3.15 & 3.12 &$0.17$ & $+0.17--0.27$\\
	& $1.0$ & $0.02$ & 3.40 & 2.66 & 2.46 & $-0.97$ & $-0.95--1.43$\\
	\enddata	
	\tablenotetext{a}{$f_{\rm ov}=0.016$}
	\tablenotetext{b}{The mixing efficiency: 100}	
\end{deluxetable*}

\subsection{Extra mixing}\label{sec:extra mixing}
\begin{figure}[t]
	\centering
	\includegraphics[scale=0.22]{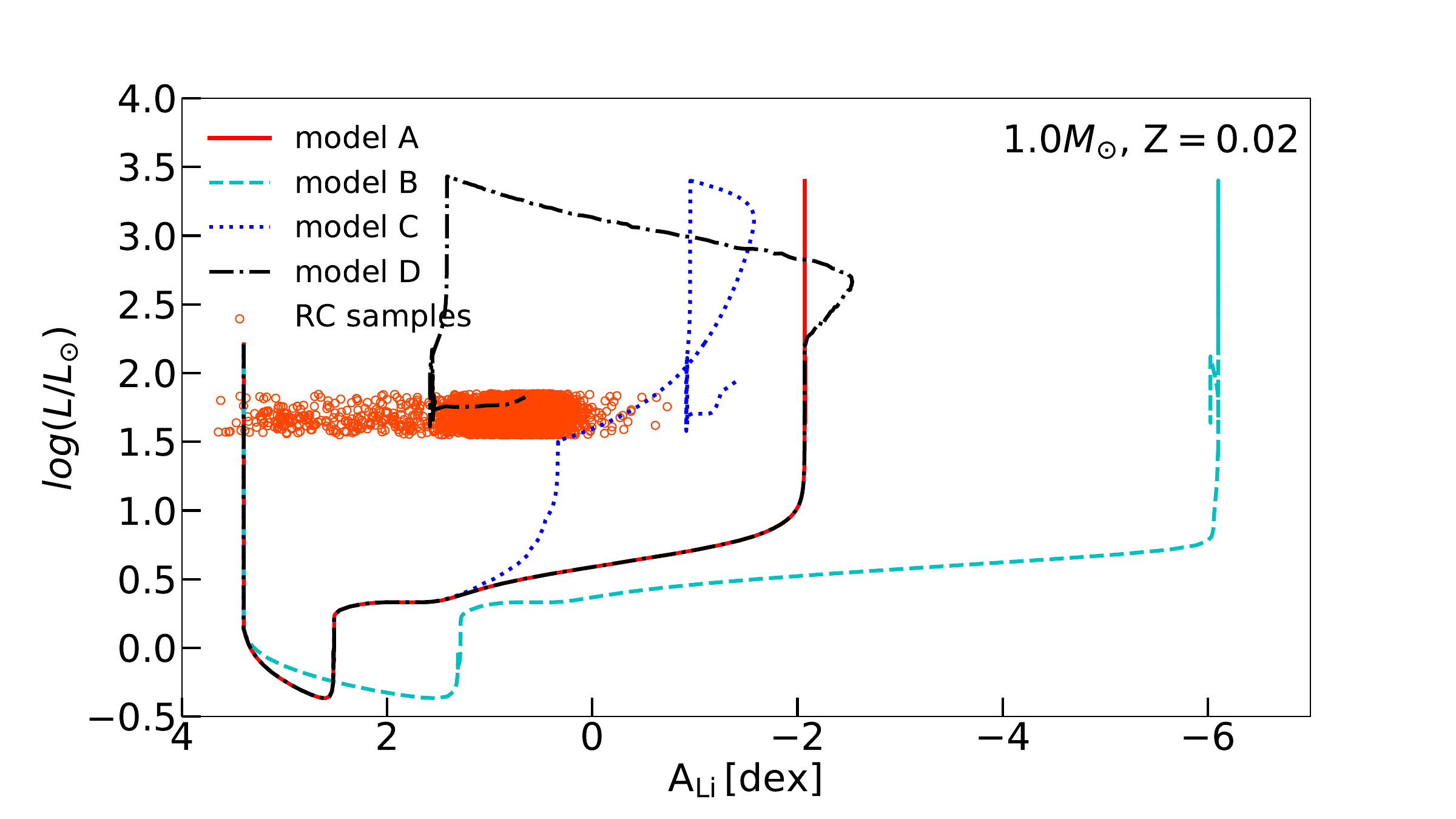}
	\caption{The evolution of Li abundance for the four different mechanisms. The four cases are as follows: \textbf{model A}: only convection, \textbf{model B}: convetion and overshooting, \textbf{model C}: convection and thermohaline mixing and \textbf{model D}: convection and internal gravity waves. The orange circles are 9,284 RC stars screened by \citet{2020NatAs...4.1059K}.
		\label{fig:f3}}
\end{figure}

For our standard convection models of solar metallicity, the mass effect cannot explain the Li abundance of RC stars. Similar situations can be found for other metallicities, which presents the shortcoming of our standard convection models. As a results, some extra mixing processes are required to investigate the Li abundances of RC stars. Overshooting can be assumed as a diffusion process, and some attempts to explain the solar Li problem with it are also underway \citep{1999A&A...347..272S, 2019ApJ...881..103Z}. The thermohaline mixing and internal gravity waves can also induce the mixing process between the H-burning shell and the convective envelope. In the following sections, we consider the impact of those processes on the Li abundances. $\rm Table\,\ref{tab:t1}$ lists the Li abundances at different stages for these models. $\rm Figure\,\ref{fig:f3}$ presents the evolution of Li abundances under the corresponding models.

\subsubsection{Overshooting}
\begin{figure}[t]
	\centering
	\includegraphics[scale=0.22]{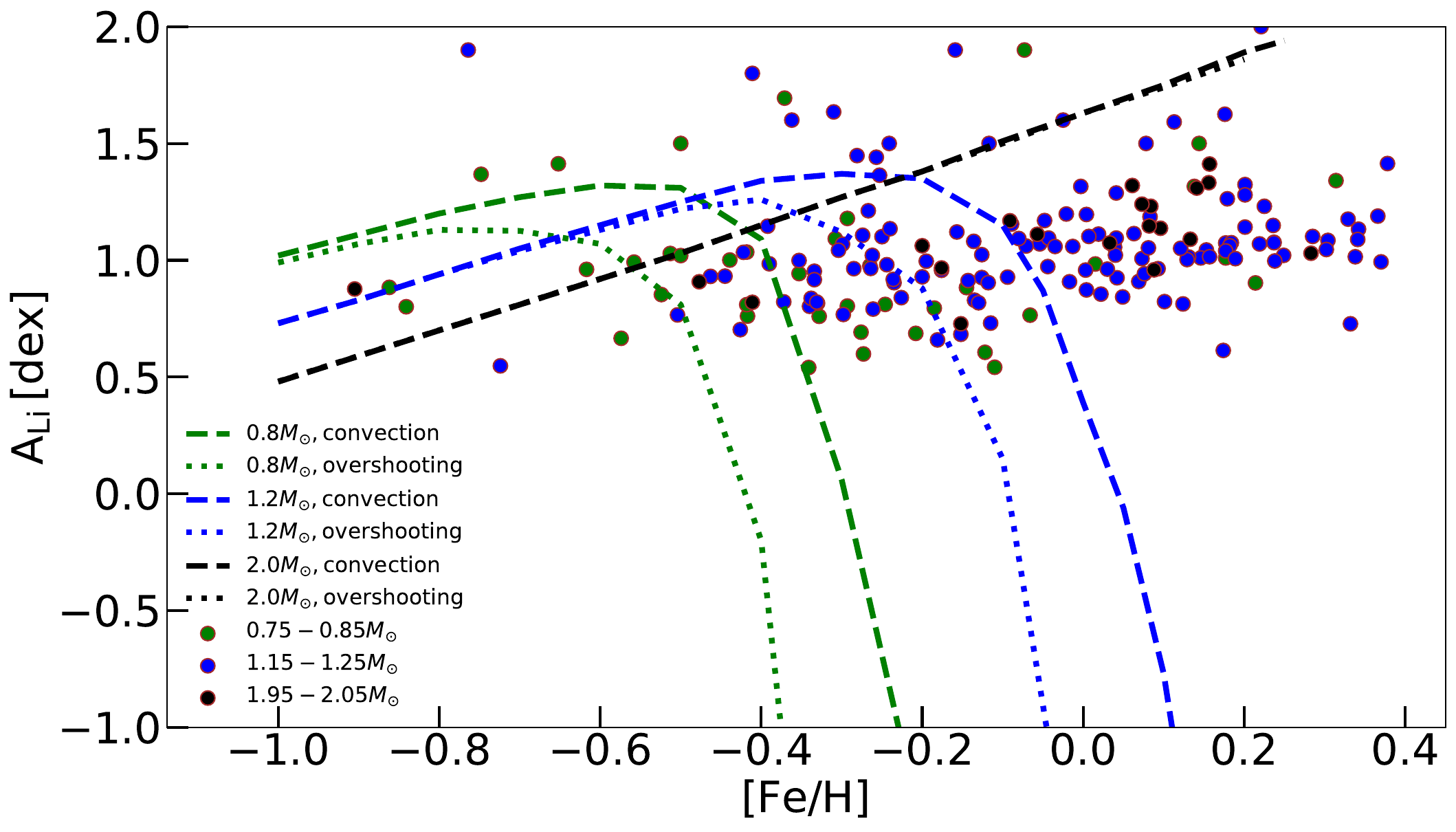}
	\caption{Comparison of overshooting and convection models. The situation is consistent with that described in $\rm Figure\, \ref{fig:f2}$. The dashed lines represent the convection models, and the dotted lines are the results of adding overshooting.}\label{fig:f4}
\end{figure}

There are overshooting at the boundaries of convective zone, and we consider the inward overshooting effect. We take the overshooting parameter $f_{\rm ov}$ as 0.016 \citep{2000A&A...360..952H}. $f_{\rm ov}$ is a free parameter characterizing the range of overshooting, i.e., extending the distance of $f_{\rm ov}H_{\rm p}$($H_{\rm p}$: pressure scale height). The larger $f_{\rm ov}$, the deeper the extension, and the more Li depletion can be achieved. \citet{2000A&A...360..952H}  reproduced the model results of \citet{1992A&AS...96..269S} when taking $f_{\rm ov}$ as 0.016. In this paper, we also adopt this typical value.

In $\rm Figure\,\ref{fig:f4}$, we compare the effects on Li abundances for the overshooting and convection models. It can be seen that when the overshooting is included, an even larger deviation can be found for stars of masses $\le1.5M_{\odot}$, while for $2.0M_{\odot}$ stars, the overshooting has basically no effect. As shown in model B of $\rm Figure\,\ref{fig:f3}$, when the overshooting is considered, the Li abundance will decay rapidly in the subgiant stages. For stars of $1.0M_{\odot}$ and $\rm [Fe/H]=0\, dex$, the Li abundance will drop to $\rm\sim -6.0\,dex$ in the subgiant stage, and it change only little in the subsequent stages. Which indicates that overshooting is not an appropriate extra mixing for the RC Li problem.

\subsubsection{The thermohaline mixing}
\begin{figure}[t]
	\centering
	\includegraphics[scale=0.22]{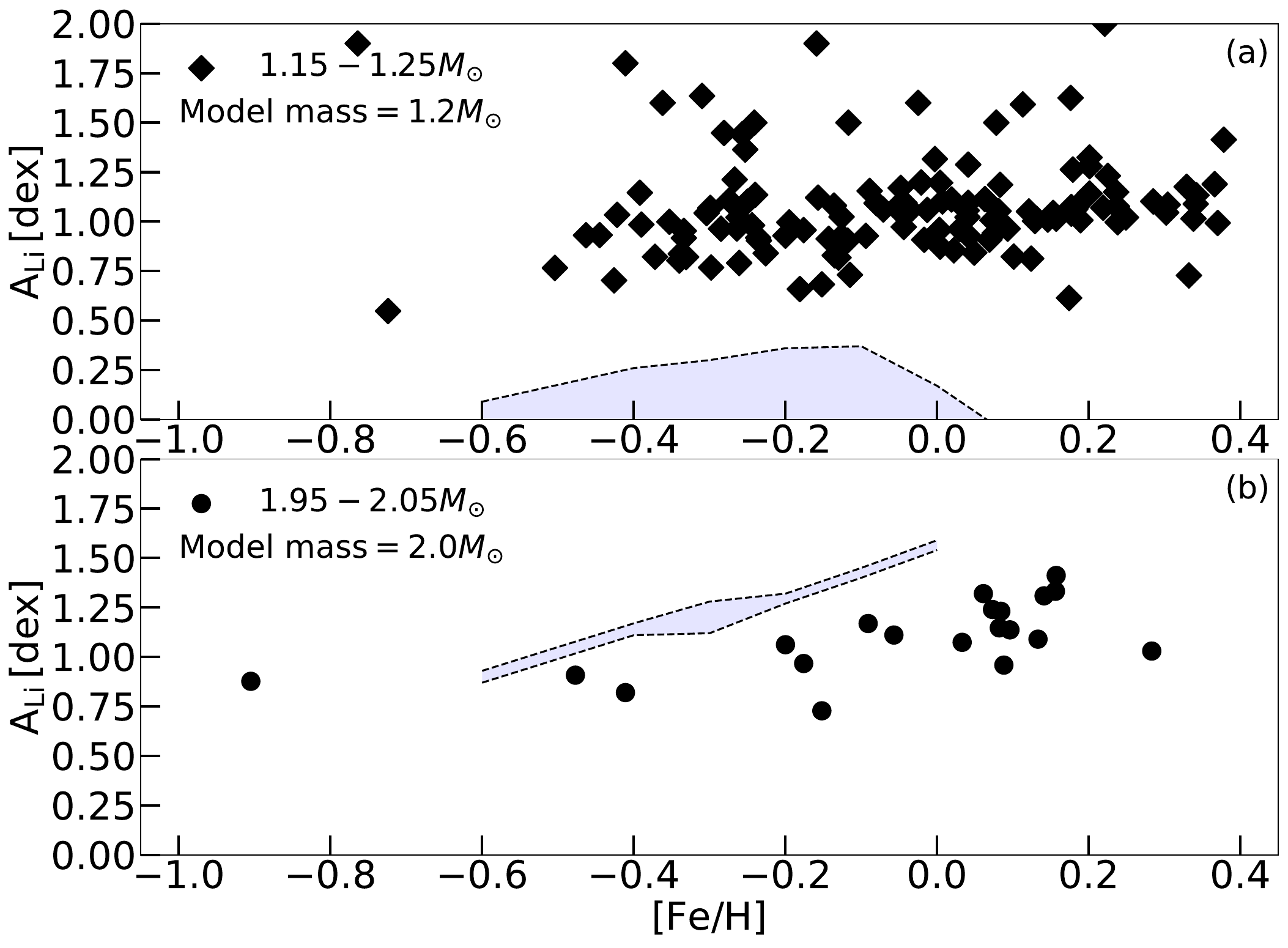}
	\caption{Comparison of the thermohaline miximg models with observations. The blue areas are model results. Panels \textbf{(a)} and \textbf{(b)} are the cases with masses of $1.2M_{\odot}$ and $2.0M_{\odot}$ respectively.}
		\label{fig:f5}
\end{figure}

In the works of \citet{2020ApJ...901L..18S} and \citet{2021MNRAS.503.2746M}, the thermohaline mixing effect is directly included in their models, so we consider its effect separately. For a solar like star, \citet{2020NatAs...4.1059K} considered the thermohaline mixing alone, and found that the model results are lower than the observed minimum Li abundance. As shown in \textbf{model C} of $\rm Figure\,\ref{fig:f3}$, at the core He-burning phase, for a typical $1.0M_{\odot}$ star, $\rm A_{Li}$ is out of the observed range. Similarly, the impact of mass and metallicity on Li abundances at the core He-burning stage is investigated.

Following the work of \citet{2020NatAs...4.1059K}, the thermohaline mixing efficiency in our thermohaline mixing models is also chosen to be 100. As shown in $\rm Table\,\ref{tab:t1}$, for the thermohaline mixing models, when the effect of mass is considered separately, the Li abundance distribution is within the range of observed data.

In $\rm Figure\,\ref{fig:f5}$, we present the Li abundance distribution of our thermohaline mixing models along with the observed samples at the core He-burning stage. When the thermohaline mixing is taken into account, the models present different Li abundance distribution characteristics from those of the convection models, and they are closer to the observed data. Nevertheless, the Li abundances still deviate from the observed range, e.g., the Li abundances of $1.2M_{\odot}$ models are lower than the observed values ($\rm Figure\,\ref{fig:f5}$ \textbf{(a)}), while it is opposite for the  case of $2.0M_{\odot}$ models. It can be seen that the thermohaline mixing models are still mainly regulated by convection.

\subsubsection{The internal gravity waves models}\label{sec:IGWS}

\begin{deluxetable*}{ccrrrrrrrrr}
\tablenum{2}
\tablecaption{The influence of coefficient $A$ on Li abundance at the core He-burning stage \label{tab:t2}}
\tablehead{\colhead{$A$} & \colhead{the mass fraction of central He} & \colhead{0.1} & \colhead{0.08} & \colhead{0.05} & \colhead{0.04} & \colhead{0.03} & \colhead{0.02} & \colhead{0.01} & \colhead{0.005} & \colhead{0.002}}
\startdata
	$\rm {A_{Li}}_{\textit{min}}$ ($\rm dex$) &0.10&$-6.51$ & $-5.99$ & $-3.33$ & $-2.31$ & $-1.21$ & $-0.14$ & $0.65$ & 0.41 & $-0.86$ \\
	$\rm {A_{Li}}_{\textit{max}}$ ($\rm dex$) &0.98& 0.52 & 0.57 & 1.00 & 1.20 & 1.45 & 1.70 & 1.55 & 0.84 & $-0.70$\\
\enddata
\end{deluxetable*}

\begin{figure*}[t]
	\centering
	\includegraphics[scale=0.25]{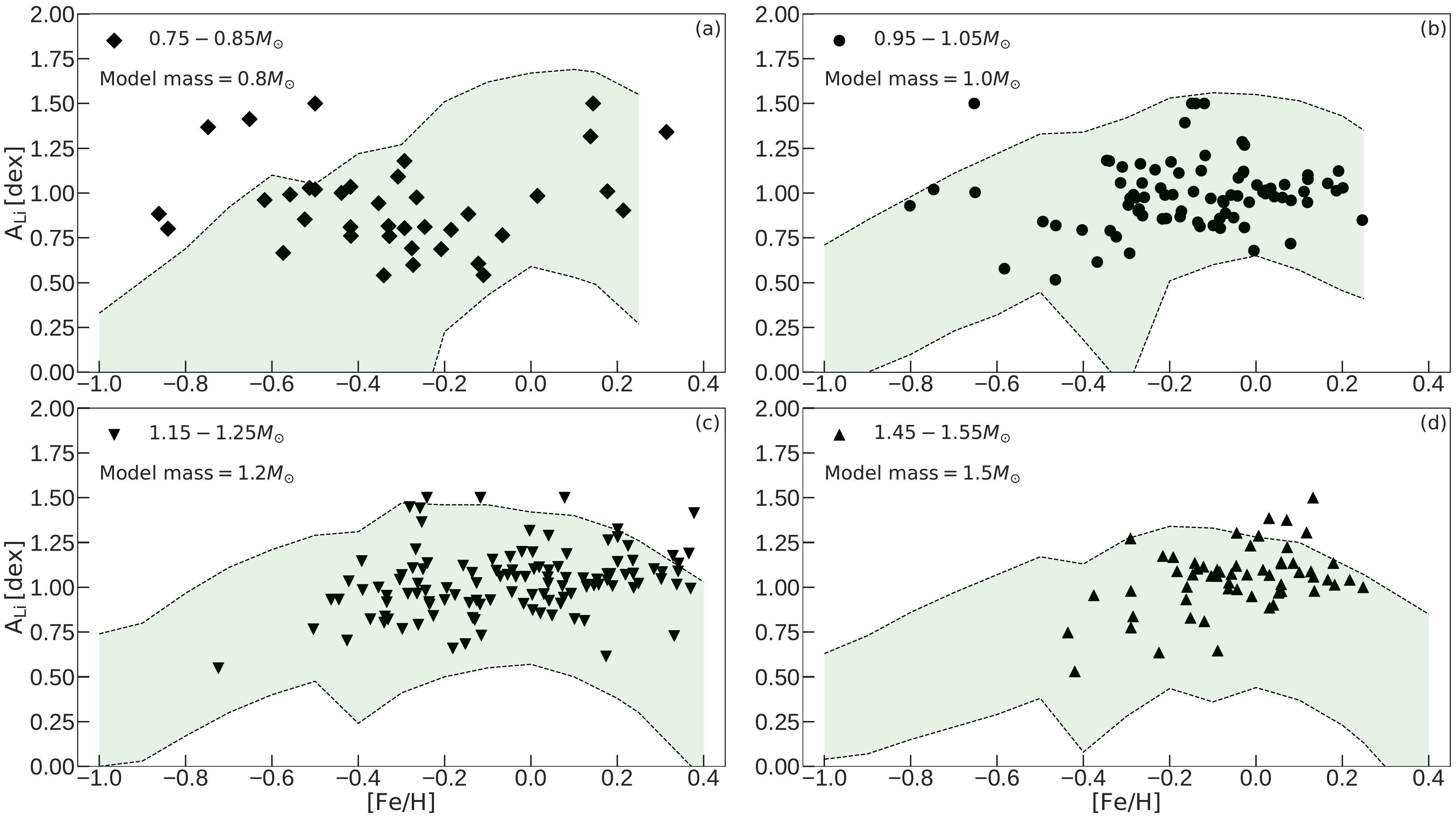}
	\caption{The distribution of observed  RC Li abundances vs those of predicted by internal gravity waves models at the core He-burning stage. The sample mass are marked with diverse shapes, and the masses of the models in the four panels are $0.8M_{\odot}$, $1.0M_{\odot}$, $1.2M_{\odot}$ and $1.5M_{\odot}$, respectively. \label{fig:f6}}
\end{figure*}

\begin{figure}[b]
	\centering
	\includegraphics[scale=0.22]{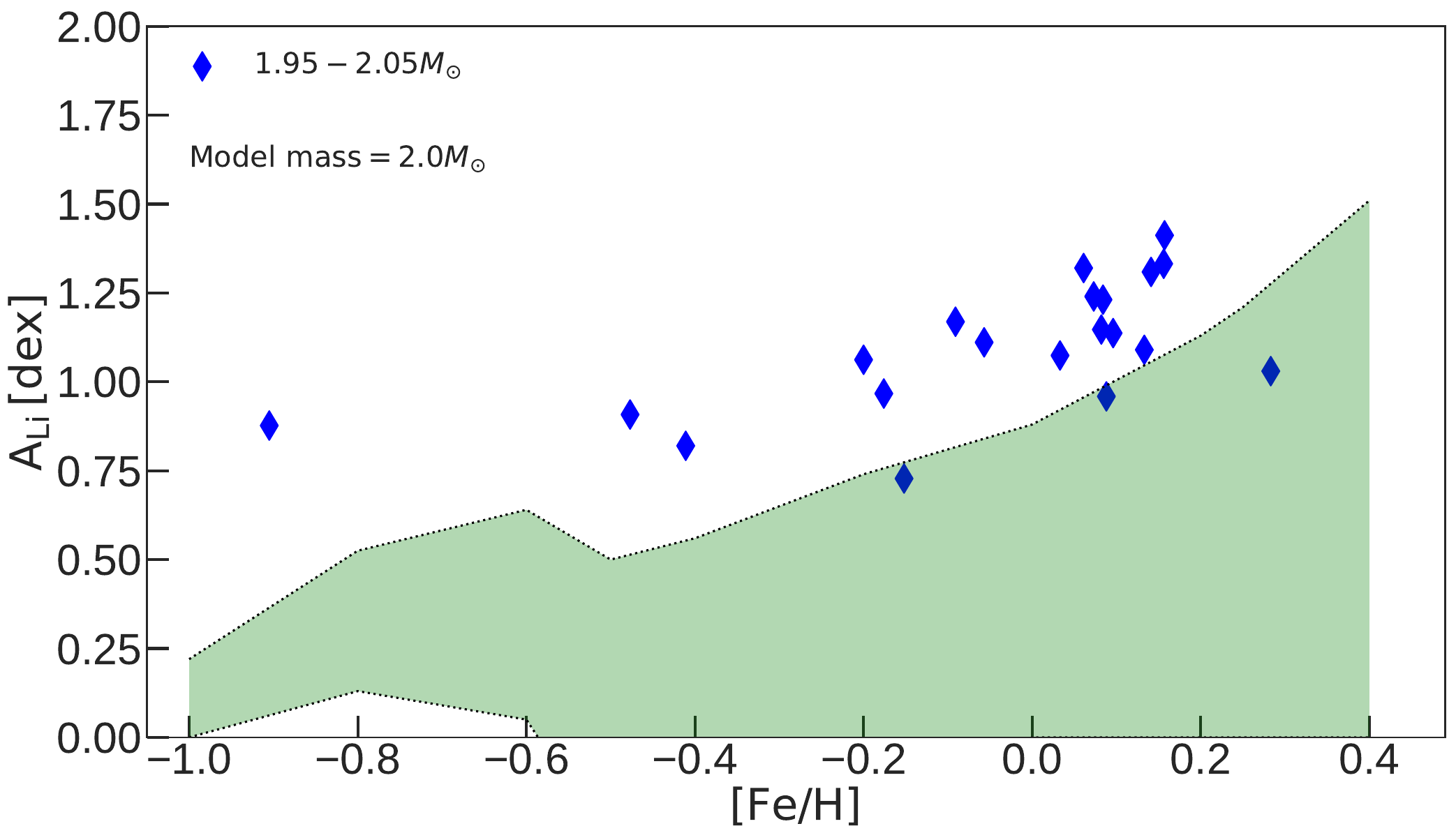}
	\caption{Similar to $\rm Figure\,\ref{fig:f6}$, but with a model mass of $2.0M_{\odot}$.} \label{fig:f7}
\end{figure}

As discussed in $\rm Section\,\ref{sec:Method}$, the internal gravity waves models are affected by the coefficient $A$. \citet{2000A&A...354..943M} carried out a 2-dimensional description of the plume perturbation. They took more number of plumes into consideration and estimated the magnitude order of $A$ at $10^{-2}\ (A \sim 0.08)$ during the MS, compared with the 1-dimensional description in \citet{1994A&A...281..421M}. However, the stellar envelope expands and the number of plumes increases during the RGB, which leads to smaller value of $A$. In order to further ascertainment, we probe the influence of diverse $A$ values on $\rm A_{\rm Li}$ during the core He-burning stage, and $\rm Table\,\ref{tab:t2}$ displays our results. When the value of coefficient $A$ is around 0.01, the distribution of $\rm A_{\rm Li}$ in the core He-burning stage is between $\sim 0.65$ and $\sim 1.55$$\rm \,dex$, which is consistent with the observed results \citep{2021ApJ...919L...3Z}. Therefore, in our models, we choose $A=0.01$.  

It needs to be pointed out that, for the internal gravity waves models, the internal gravity waves only work when the mass of He core increases to $0.3M_{\odot}$, and the Li abundances in previous stages is regulated by convection. As shown in $\rm Table\,\ref{tab:t1}$, the low-mass stars have a large degree of Li depletion during the MS as discussed by \citet{1989ApJ...347..835G}.

$\rm Figure\, \ref{fig:f6}$ displays the model results of four different masses, and the Li abundances of RC samples with corresponding mass are also shown for comparision. In these four cases, the model results cover almost all the corresponding observed Li abundance areas. Although the Li abundances of some stars are beyond the predicted range of the models, they are in good agreement with the models, as the observed results with a typical error of  $\rm 0.2\,dex$. The Li abundance range given by the models is around $\rm 0.0\,dex$ to $\rm 1.5\,dex$, which covers the normal values for most of the RC stars of \citet{2021ApJ...919L...3Z}, \citet{2020NatAs...4.1059K} and \citet{2021MNRAS.505.5340M}. However, we note that the Li abundances predicted by our models are generally lower than those of observed ones for RC stars of mass $2.0M_{\odot}$ as shown in $\rm Figure\,\ref{fig:f7}$.

\subsection{The impact of initial Li abundances on the results}
\begin{figure*}[t]
\centering
\includegraphics[scale=0.22]{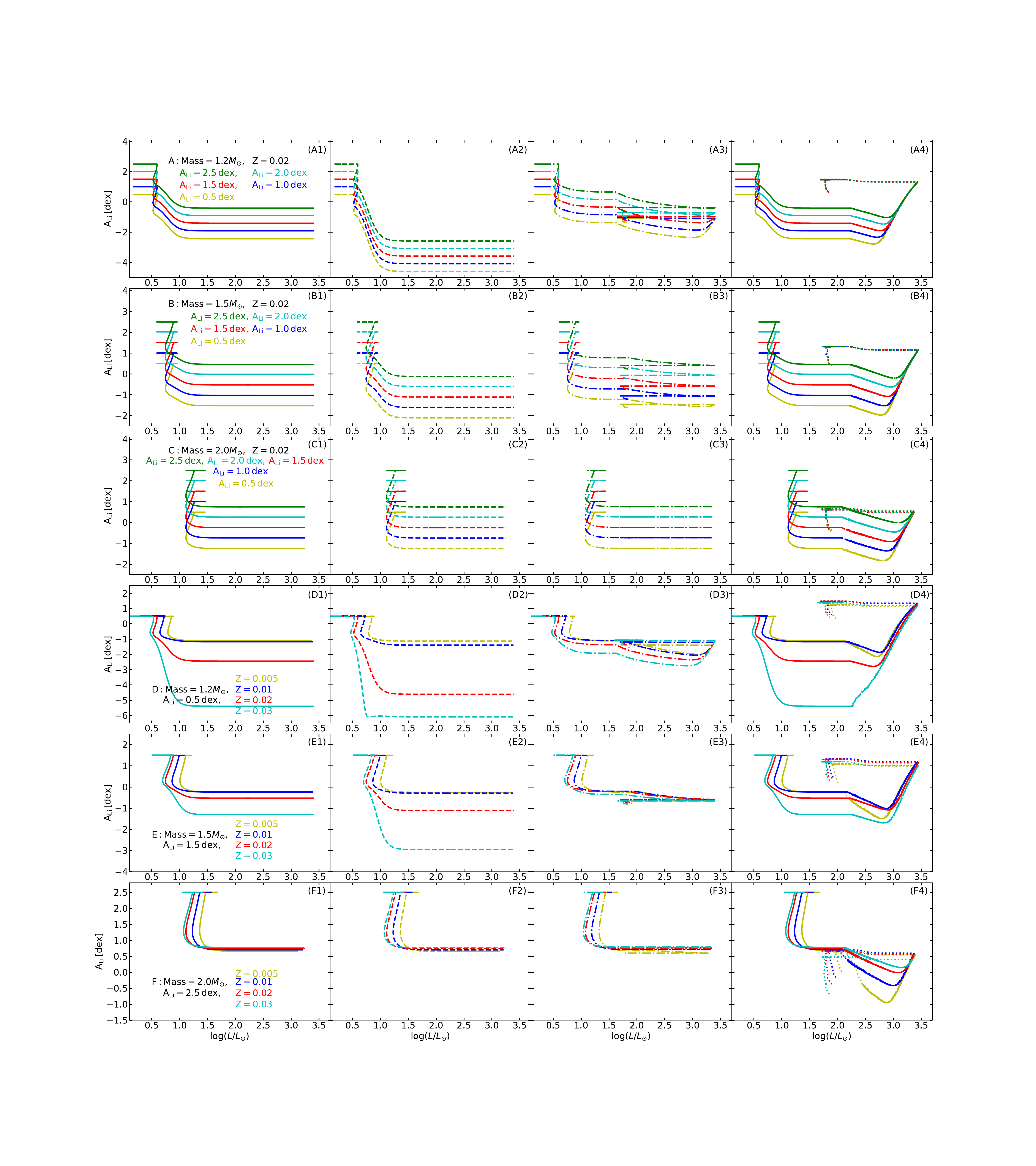}
\caption{The evolution trajectory of Li abundance with luminosity changes from the ZAMS to the end of the core He-burning. For input mass, there is \textbf{panels (A) and (D)}: $1.2M_{\odot}$, \textbf{panels (B) and (E)}: $1.5M_{\odot}$, and \textbf{panels (C) and (F)}: $2.0M_{\odot}$. Upper three panels: initial input:  Z=0.02 and the observed Li abundance distribution is used as the initial input at the ZAMS. In the lower three panels, different metallicities are used as the initial input. Number 1 (solid lines): Convection mixing only, Number 2 (dashed line): Convection + Overshooting, Number 3 (dash-dotted lines): Convection + Thermohaline mixing, and Number 4 (dotted lines): Convection + Internal gravity waves.
	\label{fig:f8}}
\end{figure*}

Before studying the Li abundance of RC stars, it is important to consider the Li depletion of the stars during the pre-MS and MS, which is exactly what \citet{2022ApJ...933...58C} did.  In $\rm Figure\,\ref{fig:f3}$, all four cases show Li depletion during the pre-MS, and there is almost no Li change during the MS in these models. However, MS stars deplete Li far in excess of standard predictions, with much evidence suggesting this is a result of rotational mixing induced by angular momentum loss \citep{1986ApJ...302L..49B, 2017AJ....153..128C, 2019AJ....158..163D, 2020ApJ...888...28B}.
Based on this, we take the observed Li abundances of stars at MS as the initial physical input to inspect the impacts of initial Li abundances on our results.

Referring to the work of \citet{2022ApJ...933...58C}, the Li abundance of MS stars with $1.2M_{\rm \odot}$ is mainly distributed between $\rm 0.0-2.5\,dex$, while around 0.5 to 2.5\,dex for 1.2 to 2.0$M_\odot$ stars.
Using the observed Li abundances as initial parameters and applying the extra mixing, we obtain the following results, as shown in $\rm Figure\,\ref{fig:f8}$. In the upper three panels of  $\rm Figure\,\ref{fig:f8}$, we take five initial Li abundances of $\rm  2.5\,dex$, $\rm  2.0\,dex$, $\rm  1.5\,dex$, $\rm  1.0\,dex$, and $\rm  0.5\,dex$ for stars with $1.2$$M_{\odot}$, $1.5$$M_{\odot}$, and $2.0$$M_{\odot}$. 
It is obviously that stars with higher mass retain more Li during the MS. As shown in the lower three panels, we take the input Li abundances of $\rm  0.5\,dex$, $\rm  1.5\,dex$, and $\rm  2.5\,dex$ for these three masses, respectively. 
In the convection-only case, the evolution of the Li abundance depends on the initial Li abundance, mass, and metallicity. For diverse initial Li abundances, the similiar degree of depletion is found in $\rm Figures\,\ref{fig:f8}$ (A1), (B1), and (C1). More Li is retained for stars with higher mass in the case of specific initial Li abundance. 
It can be noted that stars with lower mass and higher metallicity will deplete more Li (see $\rm Figures\,\ref{fig:f8}$ (D1) and (E1)). However, the Li depletion is not sensitive to metallicity for $2.0$$M_{\odot}$ star (see $\rm Figure\,\ref{fig:f8}$ (F1)).
The final RC Li abundance is similar to the result of \citet{2022ApJ...933...58C}.

The introduction of overshooting will accelerate the depletion of Li, but it only works for stars with lower mass and higher metallicity (see $\rm Figures\,\ref{fig:f8}$ (A2), (B2), (D2), and (E2)). Similar to convection, it slightly depends on metallicity for the $2.0$$M_{\odot}$ star (see $\rm Figure\,\ref{fig:f8}$ (F2)).
It can be seen from $\rm Figures\,\ref{fig:f8}$ (A3), (B3), and (C3), the thermohaline mixing is very sensitive to mass, and the higher the mass, the more Li will also be retained. However, as shown in $\rm Figures\,\ref{fig:f8}$ (D3), (E3), and (F3), the Li depletion is not sensitive to metallicity for all mass stars.

For the model of the internal gravity waves considered, the RC Li abundances do not depend on the initial Li content, while rely on the stellar mass. Similar to the results of $\rm Figure\,\ref{fig:f6}$, in $\rm Figures\,\ref{fig:f8}$ (A4) and (D4), the models with mass of 1.2 and 1.5\,$M_\odot$, the Li abundance of RC stars distribute in the observed range of $\rm 0.0-1.5\,dex$, and it does not sensitive to the differences in the MS Li content. However, the capacity of enrich Li of internal gravity waves, which are evident in the low mass stars, becomes virtually useless for 
a mass of $2.0$$M_{\odot}$ (see $\rm Figures\,\ref{fig:f8}$ (C4) and (F4)).

Although the Li abundance of most RC stars can be explained by convection, we find that internal gravity waves can also achieve this result. We note that  the Li depletion during the pre-MS and MS has little impact on the results for internal gravity waves model.

\subsection{The evolution of Li abundance}
\begin{figure*}[t]
\centering
\includegraphics[scale=0.15]{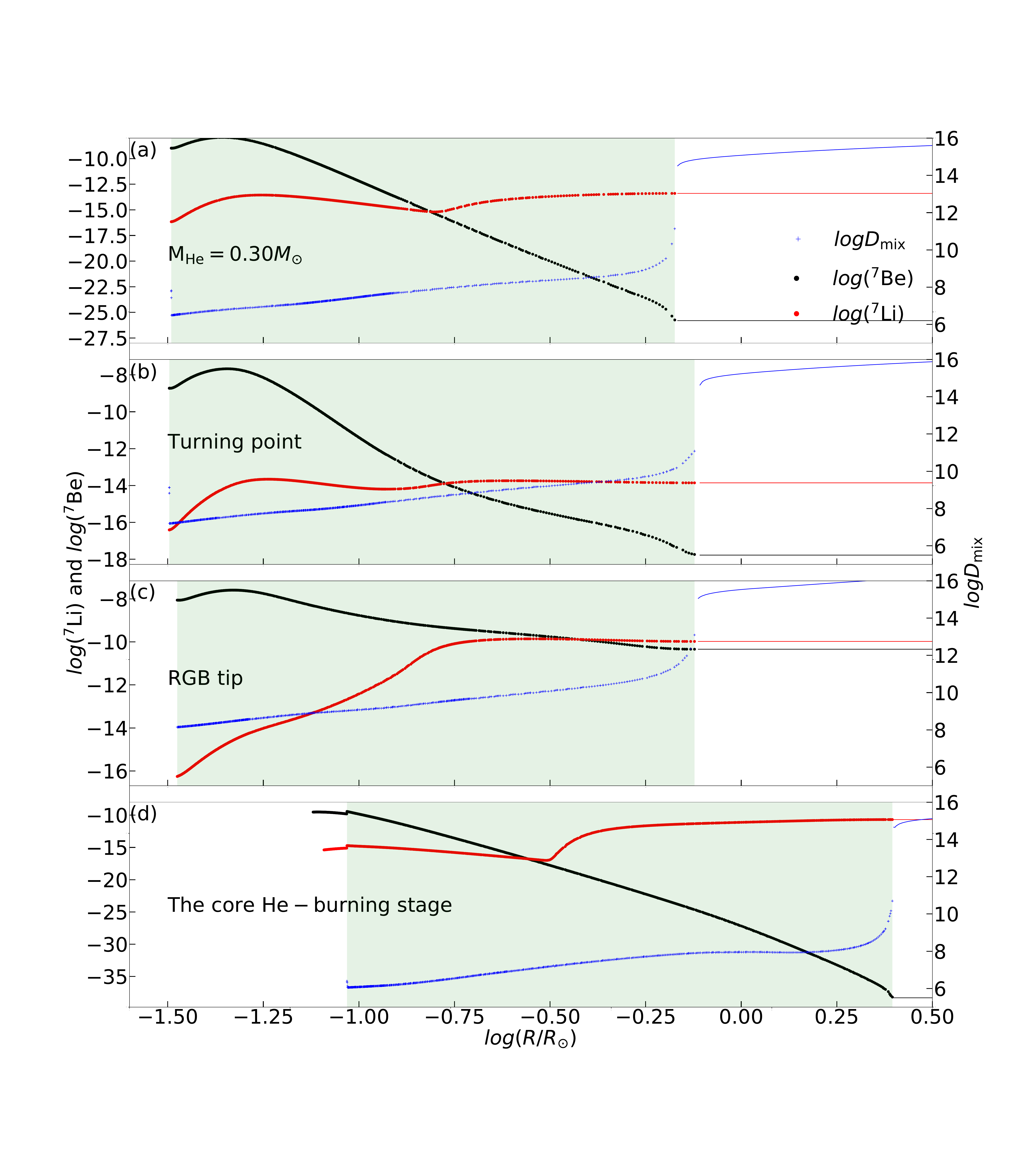}
\caption{The distribution of mass fraction of $\rm ^7Be$ and $\rm ^7Li$ and diffusion coefficient in the region of internal gravity waves action. Four nodes are selected, \textbf{(a)}: $\rm M_{He}$ = $0.30M_{\odot}$, \textbf{(b)}: the turning point, \textbf{(c)}: the RGB tip and \textbf{(d)}: the core He-burning stage. The shaded area are the mixing zone, and the two boundaries are the H-burning shell where the mass fraction H is 0.65 and the bottom of convective envelope, respectively. The left Y-axis is the mass fraction of $\rm ^7Be$ and $\rm ^7Li$, and they are represented by black and red, respectively, while the right Y-axis is the diffusion coefficient, and it is marked by blue.} \label{fig:f9}
\end{figure*}

\begin{figure}[h]
	\centering
	\includegraphics[scale=0.22]{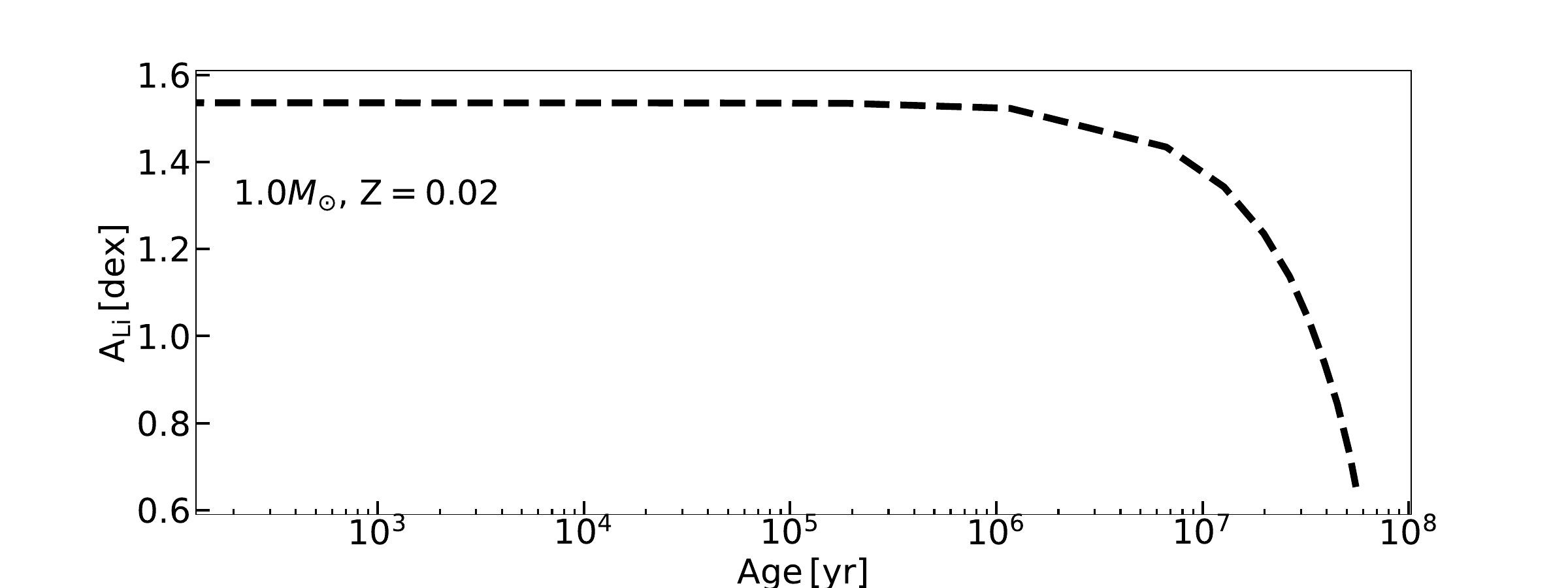}
	\caption{The evolution of Li abundance at the core He-burning stage of the internal gravity waves models. The X-axis represents the age of RC stars.} \label{fig:f10}
\end{figure}
The internal gravity waves models predict the observed Li abundances well, so we further explain how Li abundance evolves in these models.
The profiles of the diffusion coefficient and the mass fraction of $\rm ^7Be$ and $\rm ^7Li$ in the four different stages, i.e., $\rm M_{He}=$$0.3M_{\odot}$, the turning point, the RGB tip, and the core He-burning, have been described in $\rm Figure\, \ref{fig:f9}$. 

As shown in $\rm Figure\, \ref{fig:f9}$, the diffusion coefficient of mixing processes induced by internal gravity waves at the bottom of convective envelope decreases inward, and increases with the increasing luminosity and thermal diffusivity along RGB. However, at the core He-burning stage, the diffusion coefficient becomes very low relative to that of the RGB stage due to the low luminosity. The differences of diffusion coefficient at different evolution stages indicate the complexity of Li abundance evolution, as presented by \textbf{model D} in $\rm Figure\,\ref{fig:f3}$.

The evolution of Li abundance is determined by the content of Li and Be in the convective envelope, and the distribution of Be reflects the operation of Cameron-Fowler mechanism. The position, where the mass fraction of H equals to 0.65, is a little bit deeper than that of the peak Be production position at the H-burning shell, which can ensure that enough Be is transported to the envelope. From the MS to the stage where internal gravity waves can operate, the convective envelope gradually approaches the interior, and destroys Li by transporting Li from the envelope to the stellar interior with higher temperatures, therefore, the Li abundance reduces continuously. 

After the stage where internal gravity waves being activated, the mixing process makes Be produced by the H-burning shell be transported to the convective envelope continuously. With the increase of the diffusion coefficient, the content of Be in the envelope becomes higher and higher. The turning point, near $\rm log$$(L/L_{\odot})=2.7$, appears when Be transported to the envelope is in balance with Li consumed in the stars as shown in the \textbf{model D} of $\rm Figures\, \ref{fig:f3}$ and $\ref{fig:f9}$\textbf{(b)}. Before this point, the Li abundance decreases gradually, after that point, it begins to increase significantly until the RGB tip. Once the star enters the He flash phase, the rapid decline in luminosity leads to the diffusion coefficient going down quickly by orders of magnitude, which finally results in a very small increment in Li abundance. 

After entering the core He-burning stage, as shown in $\rm Figure\, \ref{fig:f9}$\textbf{(d)}, the content of Be in the convective envelope decreases to a very low level around $10^{-35}$, which is a result that mixing effect becomes weaker. Nevertheless, the diffusion coefficient is still large in the range of a little outward from the bottom of the convective envelope, so Li is carried from the envelope into the interior of the star. As a result, the Li abundance will eventually reduce. 

The evolution of Li abundance during the core He-burning is presented in $\rm Figure\, \ref{fig:f10}$. It can be seen that the attenuation of Li abundance mainly occurs in the late RC stage. In the early stage of RC stars, the core He-burning process is relatively slow, and our models show that the Li abundance does not change much, which is in line with the observed results (see their $\rm Fig\, 3.$ in \citet{2021ApJ...919L...3Z}). However, when the rapid core He-burning process occurs in the RC stars, the Li abundance will decline rapidly. During the life cycle of RC stars, our models predict a continuous Li depletion process, which is consistent with the evolutionary characteristics of RC stars.


The enrichment of Li mainly occurs from the turning point to the RGB tip under the action of internal gravity waves. However, the Li abundance dwindles constantly and finally occupies a larger range of distribution at the core He-burning stage.

The rapid enrichment of Li occurs after the turning point. As shown in $\rm Figures\,\ref{fig:f9}$\textbf{(b)} and \textbf{(c)}, the diffusion coefficient at the H-burning shell can reach as large as $\rm 10^8\,cm^{2}\,s^{-1}$. In this case, a rapid Be transfer process is induced, which will make the Li content increases from $\sim 10^{-15}$ to $\sim 10^{-10}$ during the evolution along RGB. However, when the diffusion coefficient at the H-burning shell is lower than this value, the Li abundance will decrease (see $\rm Figure\,\ref{fig:f9}$\textbf{(d)}). As a result, only when the diffusion coefficient at the location, where Be is generated, is close to $\rm 10^{8}\, cm^{2}\,s^{-1}$, the stars can enrich efficiently Li.

\subsection{The joint action of internal gravity waves and thermohaline mixing}
\begin{figure}[t]
\centering
\includegraphics[scale=0.22]{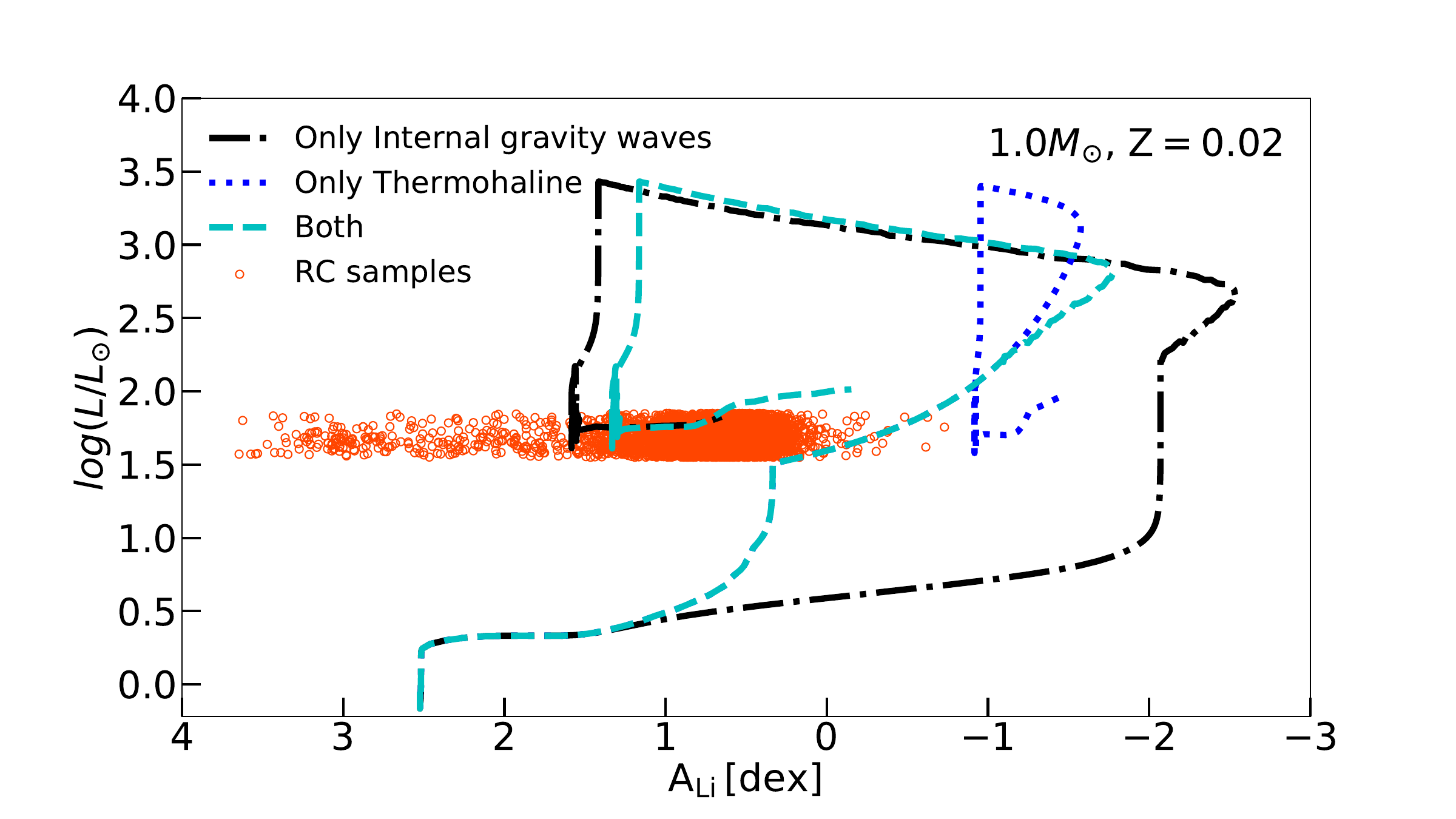}
\caption{The luminosity vs Li abundance in stars for different models. The orange circles are 9,284 RC stars screened by \citet{2020NatAs...4.1059K}.} \label{fig:f11}
\end{figure}

As shown in Figure 3 of \citet{2020NatAs...4.1059K}, the observed evolution trend of Li abundances can be explained by the thermohaline mixing in RGB stage, however, it deviates from the observations at the core He-burning stage. For the internal gravity waves models, the corresponding mixing takes into effect only when the mass of He core increases to $0.3M_{\odot}$. Before this point, the Li abundance of a star is regulated by convection. Comparing the \textbf{models C} and \textbf{D} in $\rm Figure\,\ref{fig:f3}$, we can see that the internal gravity waves models deviate from the observation in the early RGB stage. The predicted Li abundances are different from the observations on the whole evolution path when only one of the two mixing processes is considered, therefore, it is important to take both of them into account.

The evolution trajectory of Li abundance under both effects considered is shown in $\rm Figure\, \ref{fig:f11}$. It can be seen that the Li abundance during RGB stage is regulated by thermohaline mixing, and the Li enrichment due to the internal gravity waves is gradually manifested. When entering the RC stage, they form a larger distribution of around $\rm 0.0-1.5\,dex$, which is also the most densely distributed Li abundance region of \citet{2020NatAs...4.1059K}. The synergistic effects of the two mechanisms can explain the Li abundances of most giants.

\section{Discussion} \label{sec:Discussion}

In $\rm Figure\,\ref{fig:f1}$, we can find that there are very few stars with masses greater than $1.5M_{\odot}$, and they have different distribution characteristics of Li abundances from those with masses less than $1.5M_{\odot}$. The lower limit of Li abundance of such mass stars increases gradually with the mass. As shown in $\rm Figure\,\ref{fig:f7}$, for stars with $2.0M_{\odot}$, the range of predicted Li abundances is below the observations, which is exactly the opposite of the results of the other three models. For these stars, our model results show a characteristic of Li depletion, which is contrary to the above results (see $\rm Figure\,\ref{fig:f7}$) and deserves our further consideration.

At the moment, the number of these massive stars with observed Li abundance is small, thus, it is difficult to obtain a clear distribution characteristic. It is important to derive Li abundances for more such type of stars, and  to investigate this discrepancy.

\section{Conclusion} \label{sec:Conclusion}

We investigate the distribution of Li abundances in RC stars with low mass progenitors using the convection and some extra mixing models. The conclusions are as follows:


1. The internal gravity waves excited at the bottom of convective envelope are not sensitive to the depletion of Li at pre-MS and MS stages. Meanwhile, they can efficiently enhance Li, and the maximum of Li abundance can reach $\rm 1.5\,dex$. The enrichment occurs in the late RGB stage.

2. Over a relatively long evolution time at the core He-burning stage, the Li abundance predicted by our models would be almost without change, while at the end of the core He-burning stage it would decreases quickly from $\rm 1.5\,dex$ to $\rm 0.0\,dex$.



3. A significant enhancement of Li can be guaranteed only when the diffusion coeﬀicient at the H-burning shell reaches $\rm \sim 10^{8}\,cm^{2}\,s^{-1}$.  

4. The thermohaline mixing can efficiently regulate $\rm A_{\rm Li}$ in the early RGB stage, and the combination of the thermohaline mixing and internal gravity waves can explain the observed Li abundances of most giant stars.\\

We are very grateful to $\rm Prof.\, Gang\ Zhao$ for providing the RC sample of the Li abundance. We thank $\rm Dr.\,Jie\ Su$ for his guidance on the MESA code. We thank anonymous reviewers for their productive comments and fruitful suggestions. Our research is supported by National Natural Science Foundation of China (Grant Numbers: 11833006, 11973052, 11973079, 12022304, 12090040, 12090044, and 12273104) and the National Key R\&D Program of China No.2019YFA0405502. X.-F. L. acknowledges support from the Natural Science Foundation of Yunnan Province (No. 202201AT070158). H.-L. Y. acknowledges support from the Youth  Innovation Promotion Association of the CAS (id. 2019060), and NAOC Nebula Talents Program. J.-H. Z. acknowledges support from NSFC grant No.12103063 and from China Postdoctoral Science Foundation funded project (grant No. 2020M680672).

\bibliography{sample631}{}
\bibliographystyle{aasjournal}



\end{document}